\documentclass[12pt,preprint]{aastex}
\usepackage[usenames,dvips]{color}

\slugcomment{}
\shorttitle{Hydrodynamical Simulations of NGC 6782
} \shortauthors{Lien-Hsuan Lin et al.}

\begin{document}

\title{Hydrodynamical Simulations of the Barred Spiral Galaxy NGC 6782}
\author{Lien-Hsuan Lin\altaffilmark{a,b}}
\affil{Department of Physics, National Taiwan University. No 1, Sec 4, Roosevelt Road, Taipei 10617, Taiwan, R.O.C.}
\author{Chi Yuan\altaffilmark{b}}
\affil{Institute of Astronomy and Astrophysics, Academia Sinica. P.O. Box 23-141, Taipei 10617, Taiwan, R.O.C.}
\author{R. Buta\altaffilmark{c}}
\affil{Department of Physics and Astronomy, University of Alabama, Box 870324, Tuscaloosa, AL 35487,
U.S.A.}

\altaffiltext{a}{Department of Physics, National Taiwan University. No 1, Sec 4, Roosevelt Road, Taipei 10617, Taiwan, R.O.C.}
\altaffiltext{b}{Institute of Astronomy and Astrophysics, Academia Sinica.
P.O. Box 23-141, Taipei 10617, Taiwan, R.O.C.}
\altaffiltext{c}{Department of Physics and Astronomy, University of Alabama, Box 870324, Tuscaloosa, AL 35487, U.S.A.}

\begin{abstract}
NGC 6782 is an early-type barred spiral galaxy exhibiting a rich and complex morphology with multiple
ring patterns. To provide a physical understanding of its structure and kinematical properties, 
two-dimensional hydrodynamical simulations have been carried out. Numerical calculations reveal that the 
striking features in NGC 6782 can be reproduced provided that the gas flow is governed by the gravitational 
potential associated with a slowly rotating strong bar. In particular, the response of the gaseous disk to 
the bar potential leads to the excitation of spiral density waves at the inner Lindblad resonance giving 
rise to the appearance of a nearly circular nuclear ring with a pair of dust lanes. 
For a sufficiently strong bar potential, the inner 4:1 spiral density waves are also excited.  
The interaction of the higher harmonic waves with the waves 
excited at the inner Lindblad resonance and confined by the outer Lindblad resonance results in the 
observed diamond-shaped (or pointy oval) inner ring structure. The overall gas morphology and kinematical features are both well 
reproduced by the model provided that the pattern speed of the bar is $\sim 25$ km s$^{-1}$ kpc$^{-1}$. 
\end{abstract}


\section{Introduction}

It has long been recognized that a periodic potential, such as a rotating bar, can drive density waves 
in a differentially rotating disk.  A linear inviscid theory of the excitation and propagation 
of these waves in such disks was described in the seminal work of Goldreich \& Tremaine (1979, 1980).  
Subsequently, Yuan (1984) showed that the gravitational potential associated with a rotating bar can
effectively excite nonlinear density waves with high radial streaming velocities at the outer Lindblad
resonance (OLR) in the limit that self-gravity in the gas disk could be neglected. The excitation and 
wave propagation with self-gravity was later considered in the linear and nonlinear approximation by 
Yuan \& Cheng (1989, 1991).  Further progress was achieved by the development of a nonlinear asymptotic 
theory by Yuan \& Kuo (1997), who explicitly included the effects of gas pressure, viscosity, and 
self-gravitation in the formulation. However, the theory's applicability was somewhat limited as 
it was derived only in the vicinity of the resonances. Building upon this study, an improved nonlinear 
asymptotic global theory of spiral density waves was developed by Yuan \& Yang (2006) to theoretically 
examine the nearby grand-design spiral galaxy NGC 5248.

Although these nonlinear asymptotic theories can be useful in addressing the structure of disk galaxies, 
they do not provide insight on the time-dependent evolution of the structure in gas disks in response 
to a rotating bar potential.  To examine this phase, a full hydrodynamic description must be 
obtained via numerical simulations.  To this end, early numerical investigations of the evolution of disk 
galaxies driven by a rotating bar potential were carried out by Huntley et al. (1978) and Sanders 
\& Tubbs (1980).  An excellent description of the general properties of gas flows in galaxies with bars 
can be found in the investigation by Athanassoula (1992).  Simulations designed for direct comparisons 
with the observed morphology and velocity fields of specific galaxies have been published by England 
(1989) and Lindblad \& Kristen (1996) for NGC 1300 and by Lindblad et al. (1996) for NGC 1365.  

Building on this approach, we report on a quantitative comparison between observations of the barred
spiral NGC 
6782 and detailed two-dimensional hydrodynamical simulations in order 
to probe the underlying properties of the 
rotating bar. In the next section, observational data on the structure of NGC 6782 based on ground-based and HST images, and an H$\alpha$ Fabry-Perot 
velocity field, are summarized. The numerical technique for solving 
the governing hydrodynamic equations as well as the description of the initial set-up of the calculations are 
described in \S 3.  Procedures adopted for theoretically modeling the observational data are outlined in 
\S 4 and the comparisons of the numerical results with the data are presented in \S 5.  Finally, a discussion is presented in \S 6 and conclusions in \S 7. 

\section{Observational Data}

\subsection{Optical Imaging}

\begin{figure}
\figurenum{1}
\epsscale{1.0}
\plotone{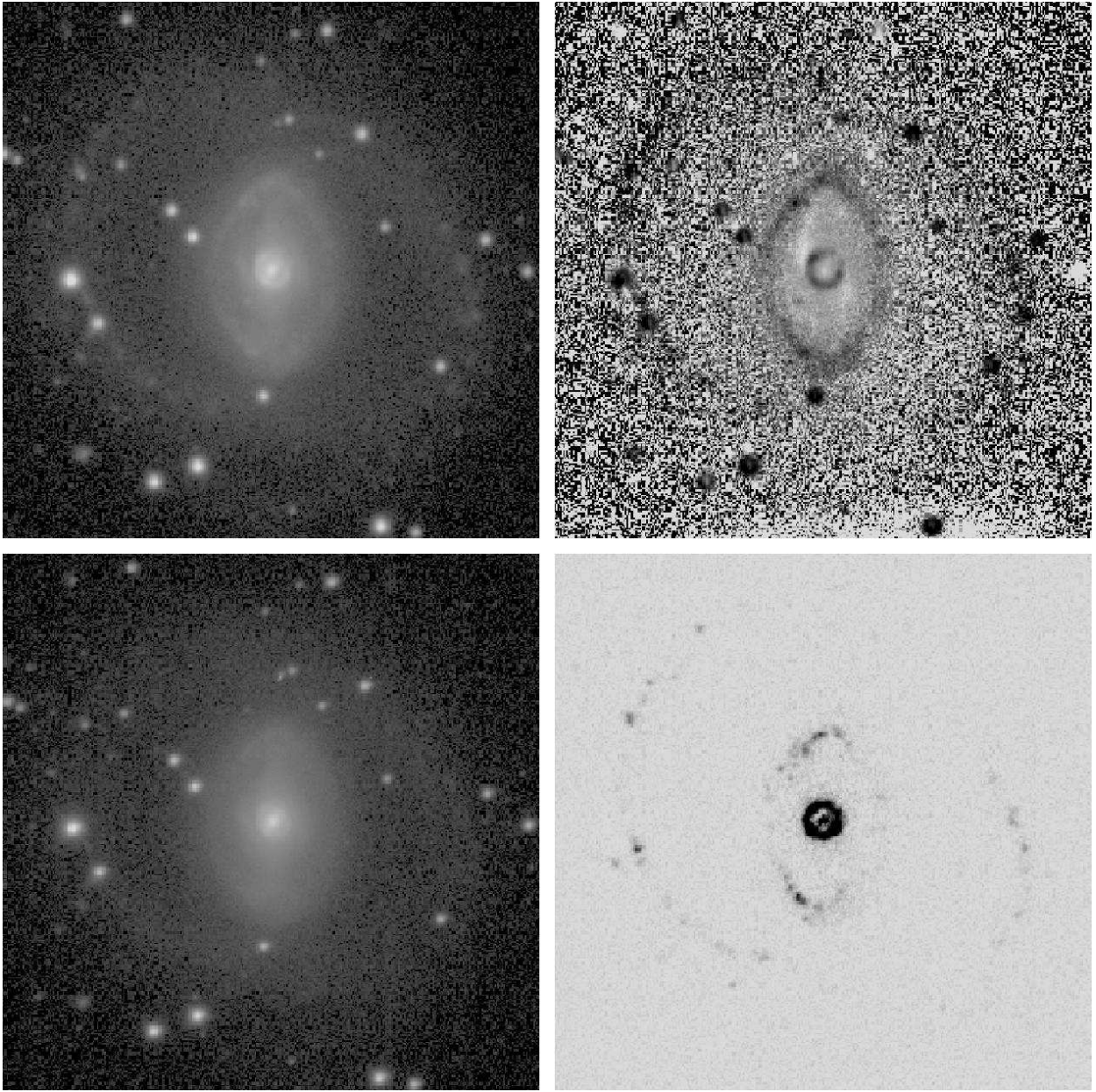}
\caption{Optical groundbased images of NGC 6782. {\it Upper left}: $B$-band image in units of mag arcsec$^{-2}$;
{\it Upper right}: $B-I$ color index map coded such that blue features ($B-I$ $\approx$ 1.0-1.5) are dark and red features ($B-I$ $\approx$ 2.5) are light.
{\it Lower left}: Red continuum image of NGC 6782 in units of mag arcsec$^{-2}$;
{\it Lower right}: H$\alpha$+[NII] image of NGC 6782 in intensity units. Both of
these latter images are from Crocker, Baugus, \& Buta (1996). North is at the top
and east is to the left in each case.}
\label{images}
\end{figure}

\begin{figure}
\figurenum{2}
\epsscale{0.75}
\plotone{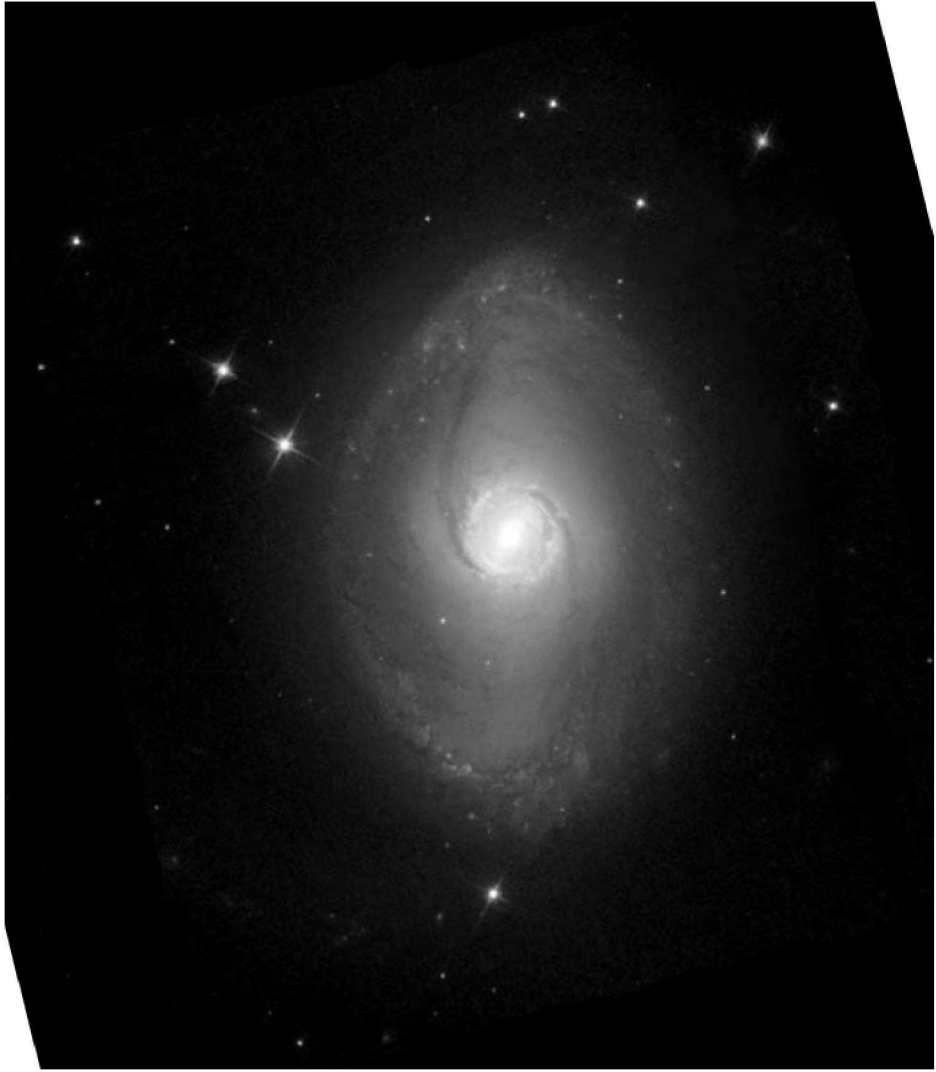}
\caption{Hubble Heritage image of NGC 6782, rotated to have north at the top and
east to the left.}
\label{hubble}
\end{figure}

NGC 6782 is an extremely interesting case, as shown by the images in Figures~\ref{images} and ~\ref{hubble}.
The type of the galaxy is (R$_1$R$_2^{'}$)SB(r)a in the system of the de Vaucouleurs
Atlas of Galaxies (Buta et al. 2007), implying a multi-ringed early-type barred spiral galaxy. 
The upper panel images were obtained with a TEK2 CCD attached to the CTIO 1.5m telescope
in 1992 August. These include a $B$-band image and a $B-I$ color index map coded such 
that blue features are dark while red features are light. A $V$-band image was also obtained,
and the exposure times were 600s in $B$, and 300s each in $V$ and $I$. The lower panel images
are 1200s exposure red continuum and H$\alpha$+[NII] images obtained in 1993 August also with 
the CTIO 1.5m 
but with a TEK1 CCD (Crocker et al. 1996). For both CCDs used, the readout noise was 
3-5$e^-$ and the gain was 
1.6$e^-$/ADU. The pixel scale was 0\rlap{.}$^{\prime\prime}$435 in each case.
All of these images were processed using standard routines in IRAF.\footnote{IRAF is
distributed by the National Optical Astronomy Observatories, which is operated by AURA, Inc.,
under contract with the National Science Foundation.}
The redshift of NGC 6782 is 0.012982$\pm$0.000137 (de Vaucouleurs et al. 1991).
We choose the value of the Hubble constant to be 100 km s$^{-1}$ Mpc$^{-1}$.
As a result, we have here adopted a distance of 38.95 Mpc, yielding a scale of 10$\arcsec$ = 1.89 kpc. 
However, the exact distance is not important in this particular investigation.

The images show that NGC 6782 has a nearly circular bright nuclear ring connected with a pair of almost straight dust lanes, 
the other ends of which attach to a diamond-shaped (or pointed-oval) inner ring.  Two faint arms in turn 
extend from the two tips of the inner ring to the outermost parts of the galaxy, forming a faint double 
outer ring-pseudoring morphology (Buta \& Crocker 1991) where the R$_1$ component is stronger. Each ring feature is a
zone of enhanced blue colors, consistent with active star formation. The pointed-oval shape of
the inner ring is an unusual feature of the galaxy and is not commonly seen. As noted by Crocker et al. (1996), the
H$\alpha$+[NII] emission is concentrated in arcs around the major axis of the inner ring, which coincides
with the bar axis. This is seen mainly in inner rings having a significant intrinsic elongation
along the bar axis. We address this concentration with our model in this paper. According to
Grouchy et al. (2008), 24\% of the total H$\alpha$+[NII] flux comes from the inner ring,
while 54\% comes from the nuclear ring and central source. The same paper gives the star formation
rate in the inner ring as 0.84 M$_{\odot}$ yr$^{-1}$. It is noteworthy that virtually all of
the recent star formation in NGC 6782 is found in its three main ring features.

Figure~\ref{hubble} shows a Hubble Heritage image of NGC 6782 that reveals finer
details of the star formation and dust in the inner regions of the galaxy. Individual
star clusters are detectable near the pointy ends of the inner ring. The weakness of
the northwest portion of the nuclear ring in the H$\alpha$+[NII] image
is due mainly to a strong dust lane. The bar
itself is very prominent in this image as a feature largely made of old stars. In addition
to the strong leading dust lanes, the image reveals very much weaker dust lanes near the
trailing sides of the bar.

\begin{figure}
\figurenum{3}
\epsscale{0.75}
\plotone{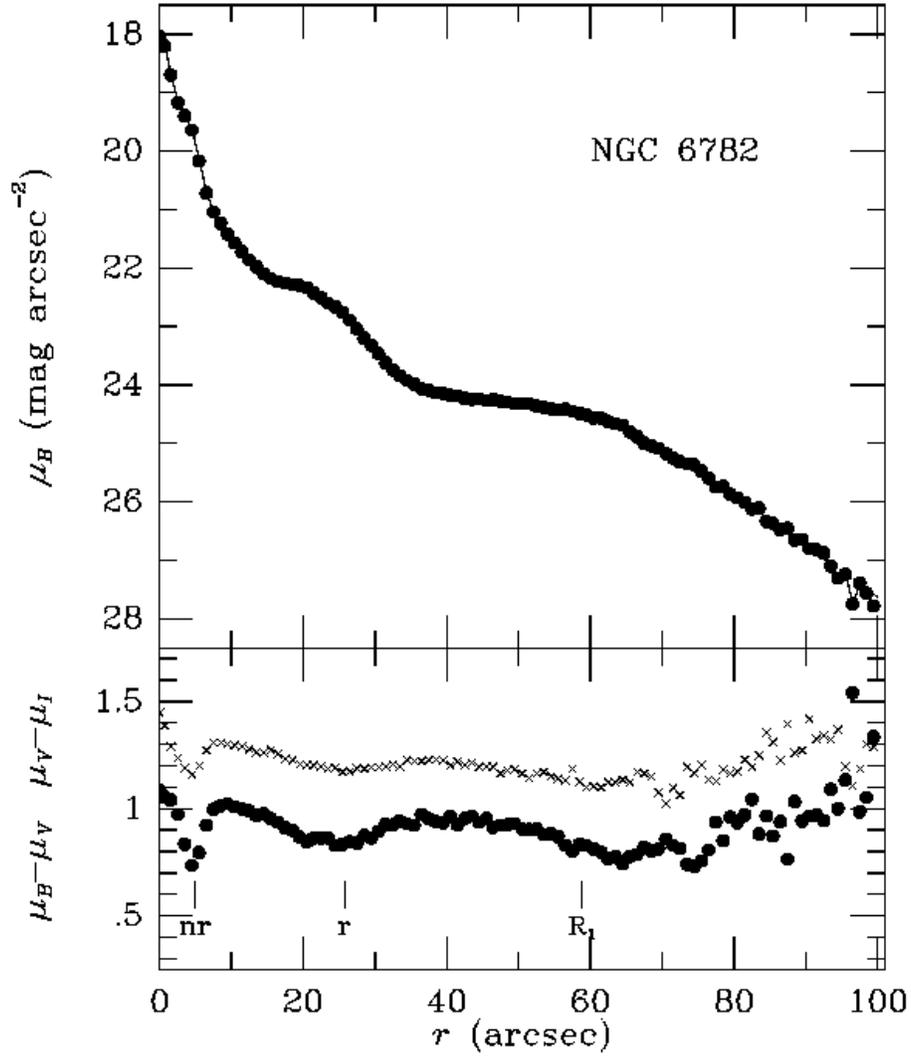}
\caption{Azimuthally-averaged surface brightness and color index profiles for NGC 6782.
The averages were taken within fixed elliptical annuli having the mean disk axis ratio
and position angle listed in Table 1. The locations of the nuclear ring (nr), inner ring (r),
and outer ring (R$_1$) are indicated.}
\label{profiles}
\end{figure}

Figure~\ref{profiles} shows azimuthally-averaged surface brightness and color index profiles of NGC 6782, based only on the CTIO images. The
surface brightness scales were calibrated from observations of Landolt (1992)
standard stars. The three rings (whose radii are from Buta \& Crocker 1993)
cause the $B$-band profile to have significant humps and the color index profiles to have
significant dips.
The $B-V$ and $V-I$ colors decrease at the location of each main ring as expected
from the color index map in Figure~\ref{images}.
These profiles are typical of early-type barred galaxies.

Table 1 summarizes some of the basic parameters of NGC 6782 derived from the 1992 broadband 
images and the profiles in Figure~\ref{profiles}. The isophotal dimensions, total magnitude,
and total colors are RC3-style (de Vaucouleurs et al. 1991) global parameters that can
improve on the previously published values. For our adopted distance, the total apparent
magnitude implies an absolute blue magnitude brighter than $-$20,
indicating that NGC 6782 is a high luminosity early-type system. 
Various published classifications indicate that
the galaxy is an SB0/a-SBa type with multiple rings. The photometric orientation
parameters are based on ellipse fits to $BVI$ outer isophotes beyond the main ring features.
These parameters are well-defined and indicate a system inclined only about 27$^{\circ}$ to the
line of sight in position angle 35$^{\circ}$. These parameters are in agreement with those 
based on analysis of the observed velocity field (section 2.2). Table 1 also includes the
deprojected diameters, axis ratios, and relative bar-ring position angles for the nuclear,
inner, and outer R$_1$ rings based on these adopted orientation parameters. The rings were
mapped visually from the $B$-band image on a TV monitor, and an ellipse was fitted to the
mappings by least squares. The inner ring
has an extreme intrinsic axis ratio (0.69) compared to the average, 0.81$\pm$0.06, found 
for SB inner rings by Buta (1995). Within the uncertainties of the visual mapping, the inner
ring is aligned almost exactly parallel to the bar. The outer R$_1$ ring is also found to
be intrinsically elongated with an axis ratio of 0.79, and is aligned almost perpendicular
to the bar. Buta (1995) found that R$_1$ rings on average have an intrinsic axis ratio of
0.74$\pm$0.08. The ratio of the outer and inner ring diameters is 2.44, somewhat larger
than the median of 2.13 found by Buta (1995) for 532 SB galaxies.
The nuclear ring deprojects into a nearly circular shape, and no bar-ring
position angle is given.

NGC 6782 is well-known as a double-barred galaxy (Buta \& Crocker 1993; Erwin 2004), with the primary bar
crossing the inner ring and the secondary bar crossing the nuclear ring. The secondary
bar is most prominent in the red continuum image in Figure~\ref{images}, and has an orientation
tipped 29$^{\circ}$ behind the primary bar (if the spiral arms are trailing). Neither
bar looks exceptionally strong, and in fact the primary bar is classified as SAB
by Corwin et al. (1985). The position angle of the primary bar is 177$^\circ$
(Crocker et al. 1996). The inner ring of NGC 6782 is so highly elongated that it also
must be considered bar-like.

\subsection{Optical Fabry-Perot Interferometry}

The optical velocity field of NGC 6782 was obtained on
1997 July 27 (UT) with the (now decommissioned)
Rutgers Fabry-Perot (RFP) interferometer
attached to the CTIO 4-m telescope.
The broad etalon with a full width at half maximum (FWHM)
of 2.3$\AA$ was used for the observations, and 
H$\alpha$ was the only emission line used for the
radial velocity measurements. The instrumental velocity
dispersion is 44 km s$^{-1}$.

Owing to the relatively low inclination,
only 16 frames were needed to cover the full velocity
range of NGC 6782. Frames ranging from 6642$\AA$ (3620 km s$^{-1}$)
to 6656$\AA$ (4260 km s$^{-1}$) were taken in 1$\AA$
steps. Each frame was exposed for 600s.
Using local position standards from the
STScI Guide Star Catalog, the scale of the frames was
measured to be 0\rlap{.}$^{\prime\prime}$349 $\pm$ 
0\rlap{.}$^{\prime\prime}$001 pix$^{-1}$.
The same stars showed that the images had to be rotated counterclockwise by
90\rlap{.}$^{\circ}$50 to get north to the top and east to the left.
Although the observations were obtained during strongly variable
cloudiness, the CTIO 4-m control system had a transparency
monitoring capability that allowed us to stop and restart
the integrations as needed when the transparency improved.
The full width at half maximum of the point spread function
on the images was 4.4 pixels (1\rlap{.}$^{\prime\prime}$5).

The main steps in the reduction of the images and the calibration of
radial velocities are described already by
Buta et al. (2001), and we refer interested readers to that paper
for details. The process gave line, continuum, radial velocity, and
velocity dispersion maps with values for more than 26,000 pixels. The H$\alpha$
line map is similar to the H$\alpha$+[NII] image shown in Figure~\ref{images}, except that
it lacks the nuclear source. This source must be dominated by [NII] emission.

\begin{figure}
\figurenum{4}
\epsscale{0.75}
\plotone{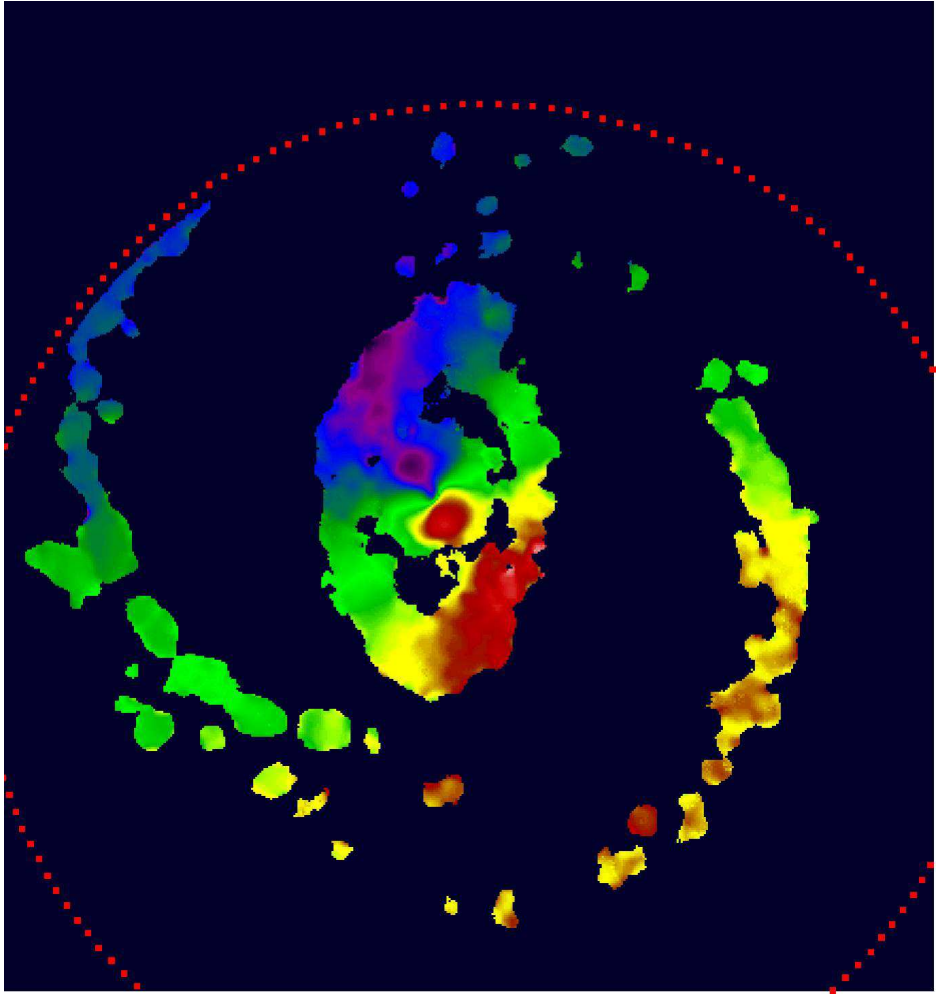}
\caption{H$\alpha$ Fabry-Perot velocity field of NGC 6782. The velocity range displayed is
from 3790 to 4060 km s$^{-1}$, and is color-coded such that receding gas is red while approaching
gas is blue. The red dotted circle shows the boundary of the instrument field and has a
diameter of 2\rlap{.}$^{\prime}$8. North is at the top and east is to the left.}
\label{vfield}
\end{figure}

The resulting heliocentric velocity field is shown with false 
colors in Figure~\ref{vfield}. The dotted partial circle indicates the boundary of the
actual field of view, which is 2\rlap{.}$^{\prime}$8 in angular
diameter. Part of the velocity field is cut-off on the northeast
side due to this boundary, but the bulk of the galaxy's ionized gas
is covered. The velocity field immediately reveals a line of nodes
inconsistent with the major axes of the inner or outer R$_1$ rings.
These features are intrinsically elongated if in the same plane, as 
is likely. At the position of the nuclear ring, two spots of high
velocity relative to the nucleus are seen. These spots are symmetrically placed around the
center and are in a position angle of 31$^{\circ}$.

Analysis of the velocity field in terms of deprojected circular motion 
(Warner et al. 1973) gave a rotation center close (within 1$^{\prime\prime}$)
to the position of the nucleus and
a systemic velocity of 3921$\pm$3 km s$^{-1}$, in excellent agreement 
with the HI Parkes All-Sky Final Survey Catalog value of 
3920$\pm$14 km s$^{-1}$ (NASA/IPAC Extragalactic Database). The
line of nodes for circular rotation was found to be in position angle 
37\rlap{.}$^{\circ}$5 by the Warner et al. (1973) method, but because
the inclination is low, the method did not strongly favor a particular
value, giving similar quality fits over a range of at least 15$^{\circ}$-30$^{\circ}$. As a check, 
we also used AIPS\footnote{Astronomical Image Processing
System, managed by NRAO, a facility of the National Science Foundation operated
under cooperative agreement by Associated Universities, Inc.} routine GAL, which
fits an exponential rotation curve to the projected velocity field.
This gave a line of nodes position angle of 32\rlap{.}$^{\circ}$9 and
an inclination of 22\rlap{.}$^{\circ}$5. This latter value is close
enough to the photometric inclination of 27$^{\circ}$ that we have
adopted 27$^{\circ}$$\pm$5$^{\circ}$ for our subsequent analysis. For the line of
nodes we adopted a position angle of 35$^{\circ}$$\pm$2$^{\circ}$, also
consistent with the photometric value.

\begin{figure}
\figurenum{5}
\epsscale{0.75}
\plotone{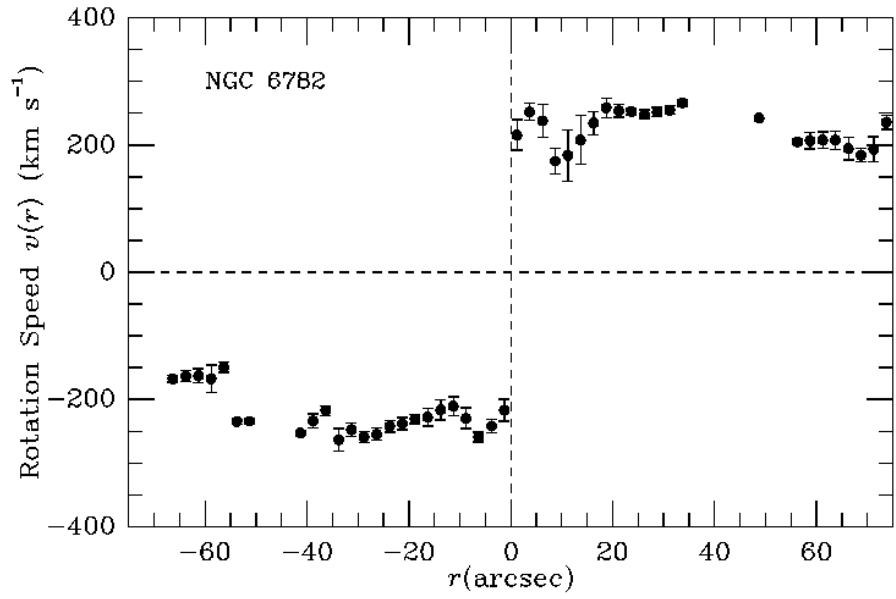}
\vspace{3.5truecm}
\caption{Unfolded rotation curve of NGC 6782. The receding half of the galaxy is
to the southwest, while the approaching half is to the northeast.}
\label{rotc}
\end{figure}

\begin{figure}
\figurenum{6}
\epsscale{0.7}
\plotone{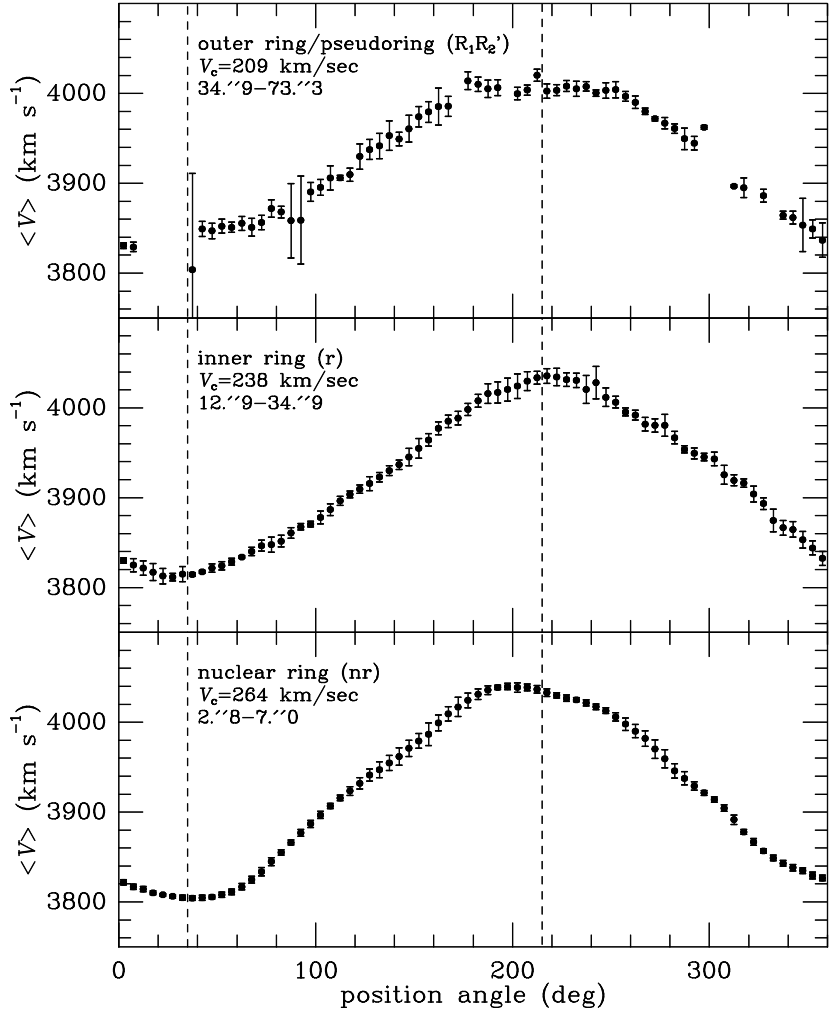}
\vspace{3.5truecm}
\caption{Mean heliocentric radial velocity versus position angle around the
three ring features of NGC 6782. The means are taken every 5$^{\circ}$ within
the indicated ranges of sky-plane radii, and the error bars are standard deviations.
The $V_c$ values are based on fits of circular rotation to these curves.
The vertical dashed lines indicate the line of nodes position angle of 35$^{\circ}$
(approaching) and 215$^{\circ}$ (receding).}
\label{vpadiagrams}
\end{figure}

Kinematic parameters are collected in Table 1. Figure~\ref{rotc} shows the 
receding and approaching halves of the rotation
curve based only on points within 30$^{\circ}$
of the line of nodes. Considerable symmetry is seen,
including a very rapid rise to a maximum rotation velocity of 
$\approx$250 km s$^{-1}$ at the radius of the nuclear ring. There
is also an indication of a slightly falling rotation curve at large
radii, again seen on both sides of the major axis. Rubin et al.
(1985) showed that the average Sa galaxy rotation curve is either flat or
slightly rising for the luminosity of NGC 6782. Since the emission
from NGC 6782 is confined to rings, two of which are intrinsically
significantly elongated, the details of the rotation curve must be 
viewed with caution. To further highlight the kinematics, we show
in Figure~\ref{vpadiagrams} mean radial velocity versus position
angle every 5$^{\circ}$ for each ring defined by the indicated
ranges of sky-plane radii. If the rotation were perfectly circular,
these curves would be symmetric, but subtle asymmetries, especially
for the inner ring, indicate the presence of noncircular motions.
Fits of circular rotation to these diagrams confirm a declining
rotation velocity, from 264 km s$^{-1}$ in the nuclear ring to 209
km s$^{-1}$ in the outer ring/pseudoring.

The outputted velocity dispersion map showed a relatively quiescent
velocity field with low dispersions throughout most of the disk.
Within the same ranges of sky-plane radii as used in Figure~\ref{vpadiagrams},
we find $<\sigma>$=33$\pm$6 km s$^{-1}$ for the nuclear ring and
11$\pm$13 km s$^{-1}$ for the inner and outer rings, where the
errors are standard deviations.
Dispersions are significantly higher in the nuclear ring,
indicating greater gas turbulence in that region.

\section{Numerical Method}

The simulations are performed with a high-order Godunov code (Antares), in which the calculation of 
hydrodynamic fluxes on zone interfaces is obtained from the exact Riemann solution (Yuan \& Yen 2005). 
Although the physical problem is naturally described using polar coordinates in cylindrical symmetry, 
there are advantages in using Cartesian coordinates in the present computations.  In particular, one 
can avoid the need to impose an artificial inner boundary condition to handle the coordinate singularity
at the origin. Moreover, the very short time steps resulting from small grid cells near the origin 
in the azimuthal direction can be avoided in using Cartesian coordinates.  At the outer boundary, 
radiation boundary conditions are imposed.  That is, wave characteristic decomposition is performed 
in each boundary cell, allowing waves to propagate outward, but suppressing incoming waves. In this 
treatment, reflection is not permitted at the boundary. 

We consider the evolution of the gaseous disk in the isothermal approximation.  The equations governing 
the flow corresponding to mass conservation and motion are given as:  
\begin{eqnarray}
\frac{\partial \sigma}{\partial t}+\nabla \cdot (\sigma \textbf{\textit{v}}) =0, \\
\frac{\partial \textbf{\textit{v}}}{\partial t}+\textbf{\textit{v}}\cdot \nabla \textbf{\textit{v}}
=-\frac{\nabla P}{\sigma}-\nabla V,
\end{eqnarray}
where $\sigma$ denotes the surface density of the disk and $\textbf{\textit{v}}$ denotes the velocity 
vector. Here, $P$ is the gas pressure integrated in the $z$-direction. For an isothermal gas,
\begin{equation}
P = a^2\sigma,
\end{equation}
where $a$ is the sound speed. The total gravitational potential, $V$, is composed of three components:
\begin{equation}
V = V_0+V_1+V_g.
\end{equation}
The first term, $V_0$, is a central potential supporting a uniform, differentially rotating disk, i.e.,
\begin{equation}
\frac{dV_0}{dr}=r\Omega^2(r),
\end{equation}
where the angular speed $\Omega(r)$ is determined from the observed rotation curve.
The second term, $V_1$, is the rotating non-axisymmetric potential of interest. For the purpose in this 
paper, we adopt a stellar bar rotating as a rigid body:
\begin{equation}
V_1(R,\phi,t)=\Phi(R)\cos[2(\phi-\Omega_pt)],
\end{equation} 
where $\Omega_p$ is the angular speed of the bar and only the first non-vanishing harmonic term is included. 
Note that the axisymmetric part of the bar potential is already contained in $V_0$.
A simplified functional form for the amplitude is adopted as:
\begin{equation}
\Phi(R)=-\Phi_0\frac{R^2}{(A^2_1+R^2)^2},
\end{equation}
where $A_1 \equiv a_1/r_s$ , $R \equiv r/r_s$, and $r_s$ = 1.0 kpc. $A_1$ and $R$ are both dimensionless.
The functional dependence on the radial coordinate, $R$, is chosen to preserve the asymptotic behavior of the potential.  It is 
proportional to $R^2$ as $R\rightarrow 0$ and approaches $R^{-2}$ for large $R$.  Thus, the force of the 
bar at $R=0$ is zero, while it behaves as $R^{-3}$ when $R$ is large. The parameter $a_1$ denotes the radial distance for which the 
bar potential is a minimum.  Finally, $V_g$, the last term of $V$, represents the self-gravitational 
potential of the gaseous disk.

The hydrodynamic code is coupled to a fast Fourier Transform (FFT) Poisson solver to include the 
self-gravitation of the disk in the calculation.  The Poisson equation for the disk is intrinsically 
3-dimensional, however, the complete potential function problem is not solved.  In contrast the force is 
calculated in the plane of the disk by integration where the force can be written as a double summation 
of the product of surface density and a kernel, which is in the form of a convolution.  The 
computational demand associated with the convolution of two vectors of length $N$ is reduced from 
$O$($N^2$) to $O$($N$ln$N$) with the FFT. As a result, the entire force computation has a demand of 
$O$($N^2$(ln$N$)$^2$).  Since the FFT is only used to accelerate the computation, our Poisson solver 
does not require the periodic boundary condition.

\subsection{Computational Setup}

The initial surface density distribution of the gas is taken to be an exponential law of the form
\begin{equation}
\sigma=\sigma_0e^{-(\frac{r}{r_0})^2},
\end{equation} 
where $\sigma_0$ is the initial surface density at the center, taken to have a value of 10 $M_\odot/pc^{2}$. 
The value of $r_0$ is specified by requiring $\sigma$ = (1/100)$\sigma_0$ at $r$ = 15 kpc. 
The actual values of the densities are unimportant, as long as the self-gravity of the gas is 
neglected, since they can be normalized without affecting Eqs. (1) and (2).  

\begin{figure}
\figurenum{7}
\epsscale{1.0}
\plotone{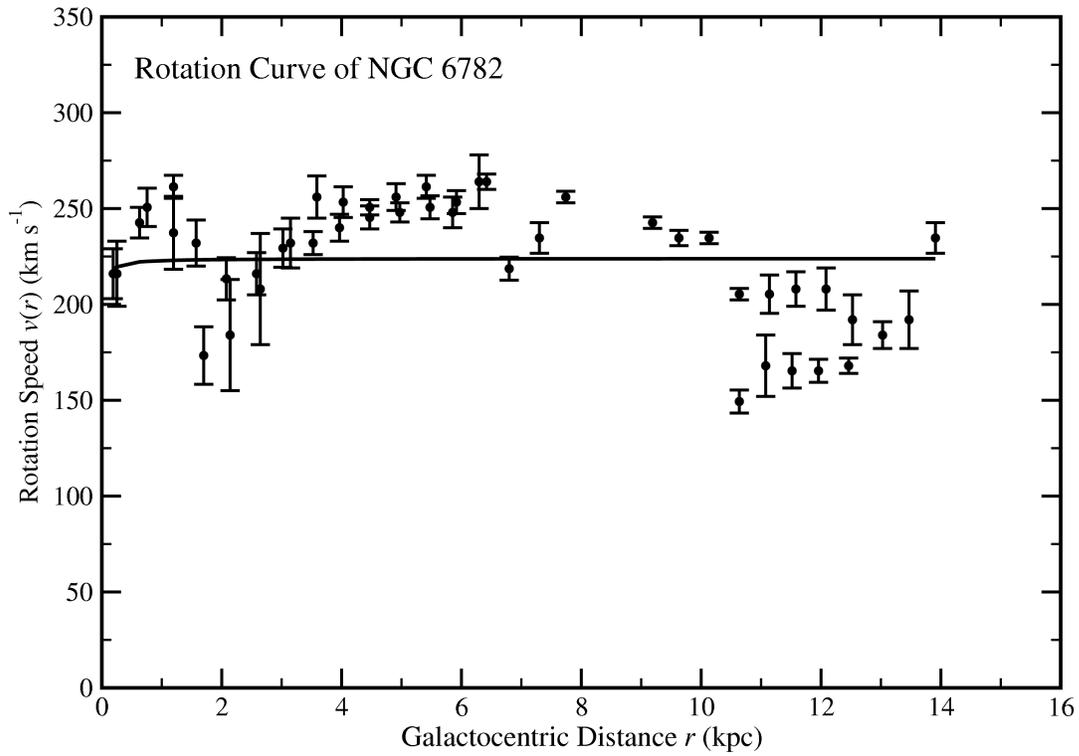}
\caption{Adopted representation of the rotation curve of NGC 6782 ($solid \  line$) for the
axisymmetric 
force in our model. The dots are the observational data points marked with their error bars, fitted 
by a nearly flat rotation curve (eq. (9)).}.\label{meanrotc}
\end{figure}

\begin{figure}
\figurenum{8}
\epsscale{1.0}
\plotone{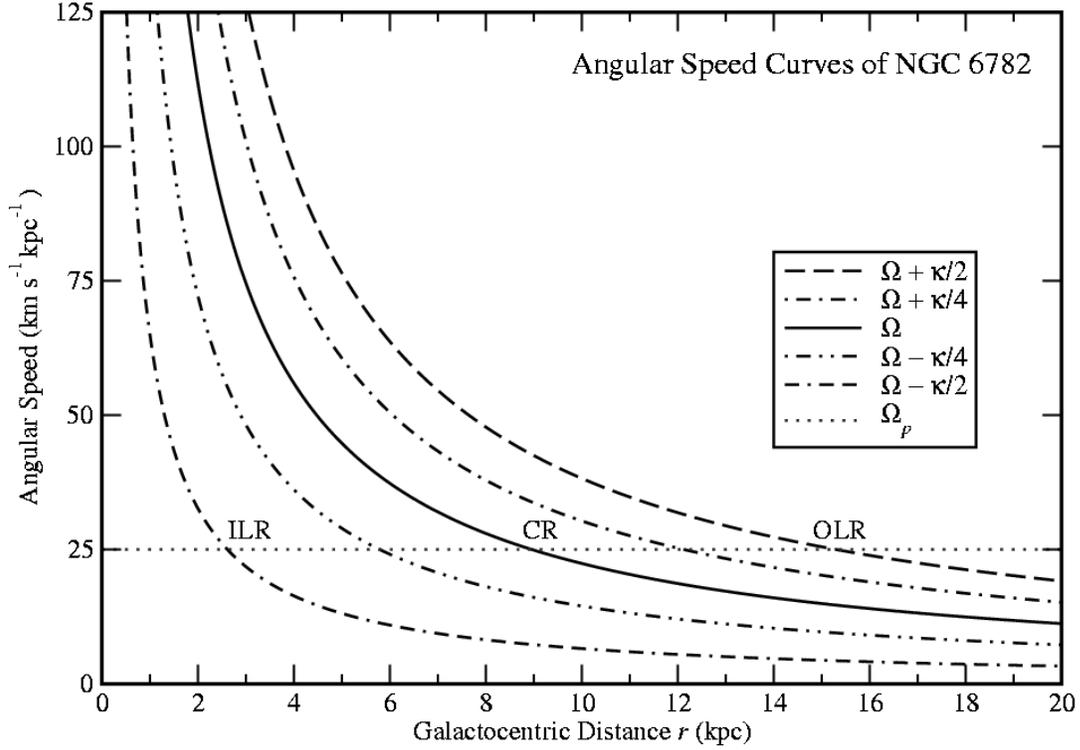}
\caption{Angular speed (precession frequency) curves as a function of radius derived from the nearly flat rotation 
curve for an angular pattern speed of the bar $\Omega_p$ = 25 km s$^{-1}$ kpc$^{-1}$. $\Omega$ and 
$\kappa$ are the circular angular speed and the radial epicyclic frequency, respectively. 
Also shown are the 
locations of corotation and the inner and outer Lindblad resonances.}.\label{pfreqs}
\end{figure}

For the axisymmetric background force, we use the representation of the rotation curve of
NGC 6782 illustrated in Figure~\ref{meanrotc}. Here the points from Figure~\ref{rotc} have
been folded around the center. Since 
it is likely that the derived rotation curve is affected by noncircular perturbations resulting from 
the spiral arms and rings, we use a least-squares approximation to fit the data points, instead of 
interpolating between them, in the form of a nearly flat rotation curve, which is written as
\begin{equation}
v(r)=v_0(\frac{r}{r+\epsilon})^{1/2},
\end{equation}
with $\epsilon$ = 0.01. The optimal value of $v_0$ is found to be 224 km s$^{-1}$. The corresponding 
precession frequency curves are shown in Figure~\ref{pfreqs}. The horizontal line represents the pattern speed of the bar, 
$\Omega_p$.  The intersection of $\Omega_p$ with the $\Omega\pm\kappa/2$ and $\Omega\pm\kappa/4$ curves, 
determine the location of the inner Lindblad resonance (ILR) and outer Lindblad resonance (OLR) as well 
as the inner and outer 4:1 resonances.  Here, $\kappa$ is the epicyclic frequency.

At the start of the simulation the gas is placed on circular orbits with rotational velocities given 
by the rotation curve.  These velocities are adjusted to take account of pressure gradients caused by 
the gradient in the density distribution to ensure that the gas is initially in a state of equilibrium. 
In order to reduce numerical noise, the amplitude of the bar potential, $\Phi_0$, is increased gradually, 
growing linearly from zero to the full value during one rotational period of the bar.  This amount of 
time was found to be necessary to avoid the formation of transients resulting from the growth of the 
perturbing potential.

To complete the specification of the input parameters, we adopt a sound speed, $a$, for the gaseous disk 
of 10 km s$^{-1}$ (Lindblad \& Kristen 1996). Provided that the sound speed is within reasonable limits, 
5$\sim$15 km s$^{-1}$, the gaseous response is not very sensitive to its value.

All the calculations presented in this paper are performed with a 1024 $\times$ 1024 uniform grid. The 
computational domain corresponds to a physical region of 40 kpc $\times$ 40 kpc, resulting in a cell size 
of 39 $\times$ 39 pc$^{2}$ (or 0.195$\arcsec \times$0.195$\arcsec$). 

\section{Modeling Procedure}
\subsection{Requirements for a Successful Model}
The appearance of NGC 6782 is bisymmetrical, and the selection criteria of a successful model 
are based on the following requirements which are similar to those used by Lindblad \& Kristen (1996):

1. The nearly circular bright nuclear ring should be reproduced.

2. The gas lanes in the model should have approximately the same shape and position as the 
observed dust lanes.
  
3. The diamond-shaped (or pointed oval) inner ring should be reproduced.

4. The faint double outer ring-pseudoring should be reproduced.

5. The characteristics in the observed velocity field, e.g. the circular rotating central disk and 
the bent isovelocity curves on the diamond-shaped ring, should be reproduced.

We judge our models by visual inspection of the model density and velocity maps overlaid on the 
observed ones instead of adopting a quantitative method, e.g. $\chi^2$, which can be misleading 
if the strong density features or steep velocity gradients present in the models do not exactly 
overlap their observed counterparts.  Such comparisons would be sufficient in the context of our 
simple model.

\subsection{Parameter Estimations}

As pointed out by Yuan \& Kuo (1997), the spiral waves excited at the outer inner Lindblad resonance 
(OILR), where the epicyclic frequency, $\kappa$, of the fluid is twice the angular frequency in the 
rotating frame of the bar, propagate inward.  The negative angular momentum carried by these waves will be 
deposited over the annular region through which they propagate.  The disk material within these regions,
after losing angular momentum, would flow toward the center.  While their theory applies only to 
the vicinity of the resonance, the results from the global nonlinear asymptotic theory of spiral density 
waves (see Yuan \& Yang 2006) reveal that the wavelength of the density waves decreases as they propagate 
inward from the OILR.  Thus, the spiral arms become more tightly wound and take the appearance of a ring 
around the center. 
We note however that these results are all based on steady-state solutions and do not describe the 
morphology after material has been swept by the waves and accreted to the center.  By carrying out 
numerical simulation experiments, we find the quasi-steady state of a bar-driven gaseous disk with 
an OILR is characterized by an "eye-shape" morphology which includes a pair of open, strongly shocked 
spiral arms in the outer depleted region and a pair of tightly wound spiral arms embedded in a more 
widespread ring in the circumnuclear region. Based on these experimental results, an OILR is required 
in order to simulate the nuclear ring and the gas dust lanes, confirming the results obtained by 
Athanassoula (1992).

An additional test used to validate our numerical simulations is to consider the case when the OLR and OILR simultaneously 
exist.  For this test, a proper angular speed for the bar is chosen such that the locations of the 
OLR and the OILR are in the appropriate range.  If the strength of the bar is weak, 
waves excited at the OLR propagate outward and those at the OILR propagate inward and no waves are 
excited in between, as predicted by linear theory.  By increasing the strength of the bar, the 
evanescent (wave-forbidden) region between the OLR and OILR narrows in extent and the inner 4:1 
higher harmonic spiral density waves between them are excited.  Their interaction with the waves 
excited at the OILR form a shape similar to the diamond-shaped inner ring seen in NGC 6782.  Based 
on this test and the observations, the values for the angular pattern speed, the location of the 
potential minimum, and the strength of the bar potential can be obtained. 

\subsection{Parameter Space}

The pattern speed $\Omega_p\,$ controls the positions of the resonances and thus the morphology of 
the spirals.  According to the images of NGC 6782 and the results of our simulation tests, $\Omega_p$ 
takes values between 25 $\leq \Omega_p \leq$ 38 km s$^{-1}$ kpc$^{-1}$, corresponding to a radial 
interval for the radius of the OLR of 10 kpc $\leq$ R$_{OLR}$ $\leq$ 15.3 kpc.

The location of the minimum and the strength of the bar potential can also be determined.  In particular, 
our test results show that the nuclear ring forms inside the location of the minimum and that the 
stronger the bar potential, the smaller the ring.  Considering both factors simultaneously, the 
location of the minimum of the bar potential, a$_1$ in Eq.7, is taken in the range 2.3 $\leq$ a$_1 
\leq$ 3.2 kpc.  To parameterize the strength of the bar potential, we denote by B the ratio of the 
effective radial force exerted on the disk by the rotating bar (Yuan \& Kuo 1997) to the force for 
the circular motion from the rotation curve at the location of OLR.  This choice is motivated by the 
fact that the waves are excited at the resonances.  It is found that B should lie between 3\% and 10\%. 

\section{Numerical Results}
\subsection{Comparisons with the Observations}
In order to quantitatively compare our theoretical results with the observations, the plane of the 
theoretical disk must be oriented with respect to the line of sight. From the analysis of the 2D 
Fabry-Perot velocity field, the P.A. of the line of nodes and the inclination of NGC 6782 are 
35$^\circ$ and 27$^\circ$ respectively. The P.A. of the major bar of the galaxy is 177$^\circ$.  

\begin{figure}
\figurenum{9}
\epsscale{1.0}
\plotone{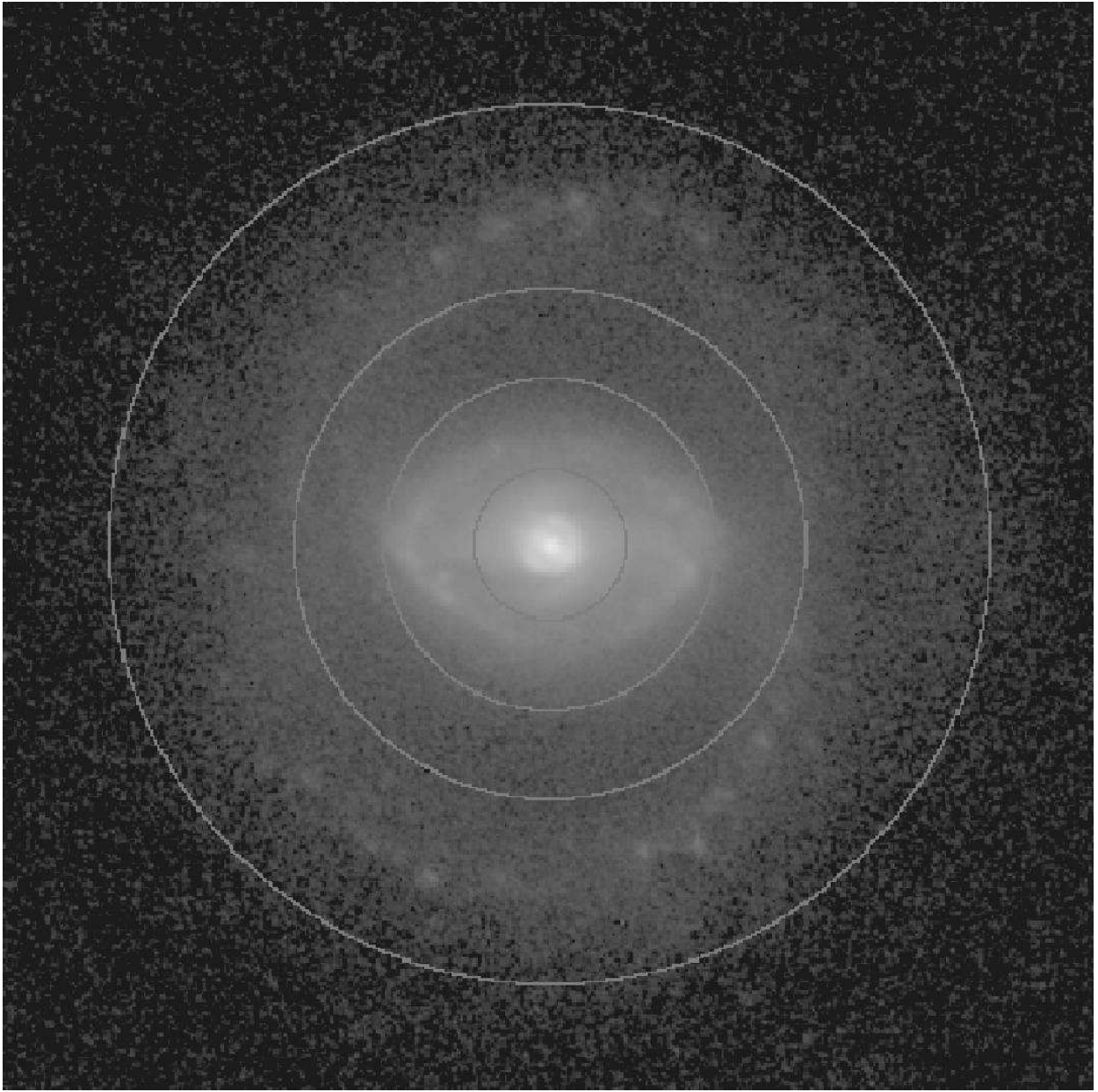}
\caption{Locations of resonances in NGC 6782 superposed on a deprojected $B$-band
image rotated such that the bar axis is horizontal. In order of increasing radius, the circles
are ILR, UHR (inner 4:1), CR, and OLR.}
\label{resonances}
\end{figure}

\begin{figure}
\figurenum{10}
\epsscale{0.8}
\plotone{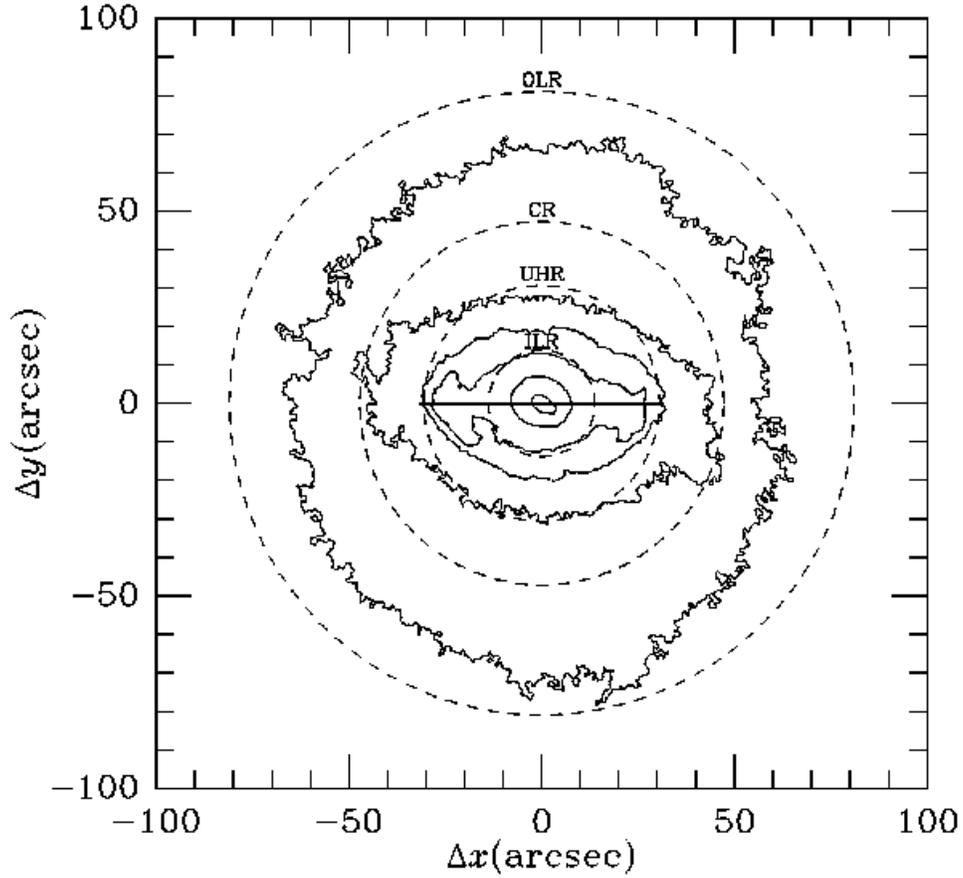}
\vspace{3.5truecm}
\caption{The same resonance radii ({\it dashed circles}) as in Figure~\ref{resonances} superposed on
a contour map of $B$-band isophotes at 19.2, 21.0 22.2, 22.8, 24.0, and 25.00 mag arcsec$^{-2}$.
The solid horizontal line indicates the extent of the bar estimated by Rautiainen et al. (2005).
This plot shows that CR in our model lies well beyond the bar ends, but fully encompasses an
extended oval only slightly misaligned with the main bar.}
\label{resonances1}
\end{figure}

The best-fit values of the parameters to the observations are $\Omega_p$ = 25 km s$^{-1}$ kpc$^{-1}$,  
a$_1$ = 3.2 kpc, and B = 5\%.  The inferred value of $\Omega_p$ implies the positions of the 
inner/outer Lindblad resonances (ILR/OLR), the inner 4:1 ultraharmonic 
resonance (UHR), and the corotation resonance (CR)
are located at R$_{ILR}$ = 2.61 kpc, 
R$_{UHR}$ = 5.78 kpc, R$_{CR}$ = 8.92 kpc, and R$_{OLR}$ = 15.3 kpc. These
locations are shown relative to the galaxy's morphology in Figures~\ref{resonances} and ~\ref{resonances1}. In Figure~\ref{resonances}, the circles are superposed on a 
deprojected $B$-band image where the bar axis has been rotated to a horizontal
orientation. The inner ring is confined between ILR and UHR, 
while the outer R$_1$ ring lies between CR and OLR,
close to
the {\it outer} 4:1 UHR at R = 12.1 kpc (where $\Omega_p=\Omega+\kappa/4$)
according to Figure 8. The positioning of the R$_1$ ring is very
similar to what Treuthardt et al. (2008) found for the
barred spiral galaxy NGC 1433 using sticky particle
simulations. 
The nuclear ring is
well inside ILR. Figure~\ref{resonances1} shows the same circles superposed
on a contour plot of six $B$-band isophotes. In this plot, the horizontal
line indicates the extent of the bar as estimated by Rautiainen et al. 
(2005). The plot shows that the isophotes of NGC 6782 continue to be
oval and nearly aligned with the bar all the way to CR. 

\begin{figure}
\figurenum{11}
\epsscale{1.0}
\plotone{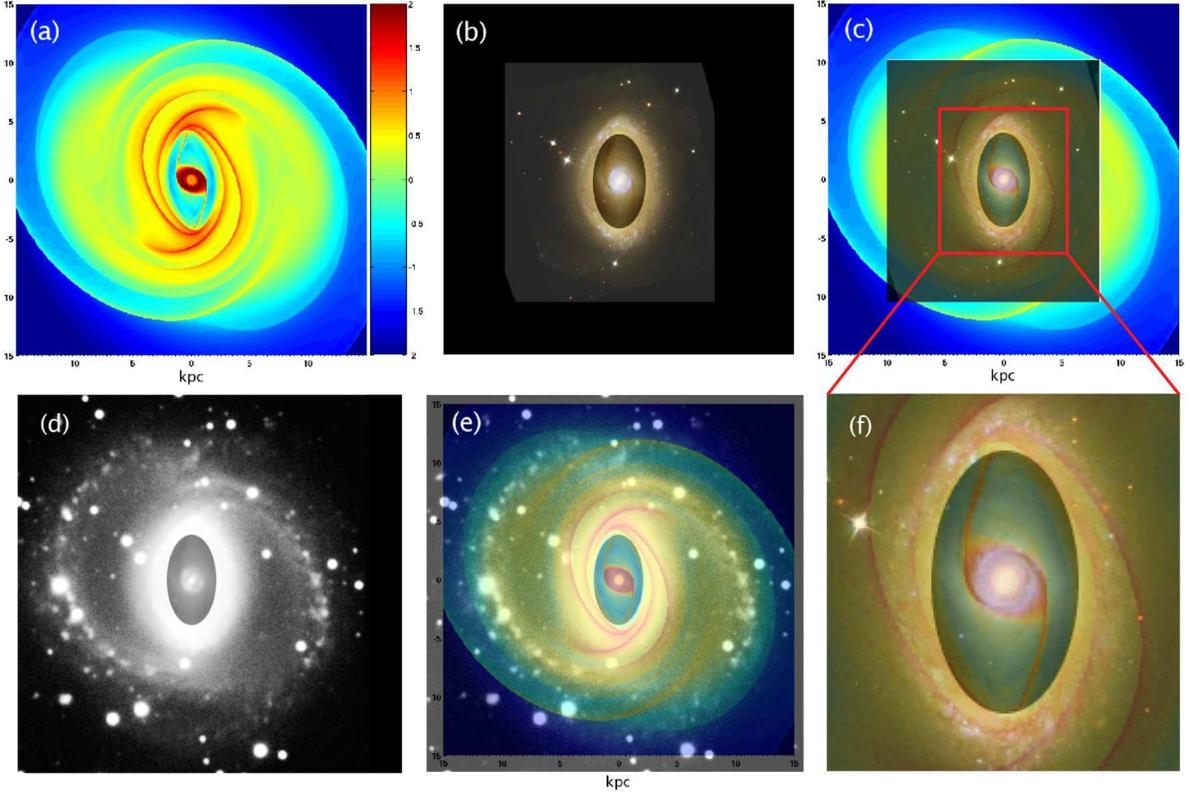}
\caption{(a) The projected density distribution of the simulation result. The color map denotes the surface
density distribution in logarithmic scale. The frame is
taken after 2.16 revolutions of the bar, or 532 Myr. 
(b) Replica of Figure~\ref{hubble} with the brightness of the pointed-oval inner ring enhanced.
(c) Superposition of Figure~\ref{surfdens2}a and Figure~\ref{surfdens2}b. The bright nuclear ring, off-centered gas lanes, as well
as the pointy oval inner ring in our model exhibit the same shapes and positions as the observation.
(d) Replica of the $B$-band image in Figure~\ref{images} with the brightness enhanced except for the central part. 
(e) Superposition of Figure~\ref{surfdens2}a and Figure~\ref{surfdens2}d. The two faint branches near the tip of the inner ring 
at the northwestern outer arm in the $B$-band image are also reproduced in our simulation result.
(f) A close-up of the inner regions of Figure~\ref{surfdens2}c. Except Figure~\ref{surfdens2}f, all panels in this figure are of the same scale.}\label{surfdens2}
\end{figure}

The projected surface density distribution for the best-fit is illustrated in Figure~\ref{surfdens2}a.  Its 
superimposition onto the optical multicolor image and the $B$-band image are shown in Figures~\ref{surfdens2}c and 
~\ref{surfdens2}e, respectively. Specifically, Figure~\ref{surfdens2}c shows that the location and shape of the bright nuclear ring, 
the dust lanes, as well as the pointy oval inner ring are reproduced by the simulations. 
At the northwestern outer arm in Figure~\ref{images}, upper left panel, there are two 
faint branches near the tip of the inner 
ring, which are also reproduced by our simulation (see Figure~\ref{surfdens2}e).
We note, however, that the fit is not without deviations since the simulation cannot fit the 
outer part of the outer arms.  Observationally, the arms extend almost one half of a revolution, but 
those in our simulation only extend one fourth revolution and are much broader.
This is because only the gas response is considered in the simulation and the stellar response has not been taken into account.
Although it is possible to produce longer spiral arms in the simulation by 
adding a weak spiral potential (Lindblad et al 1996, Lindblad \& Kristen 1996), we have chosen 
not to do so in this paper.  The imposition of a spiral potential can generate  waves by the 
resonant excitation as well as by forced oscillation (Wang 2006).  These two types of waves may 
interact, thereby complicating the morphology.  Since the main features in the images of NGC 6782 are 
well delineated and smooth, we seek to use a simple model to reproduce these features as much as 
possible in order to gain a fundamental understanding of their formation and evolution. 

\begin{figure}
\figurenum{12}
\epsscale{0.6}
\plotone{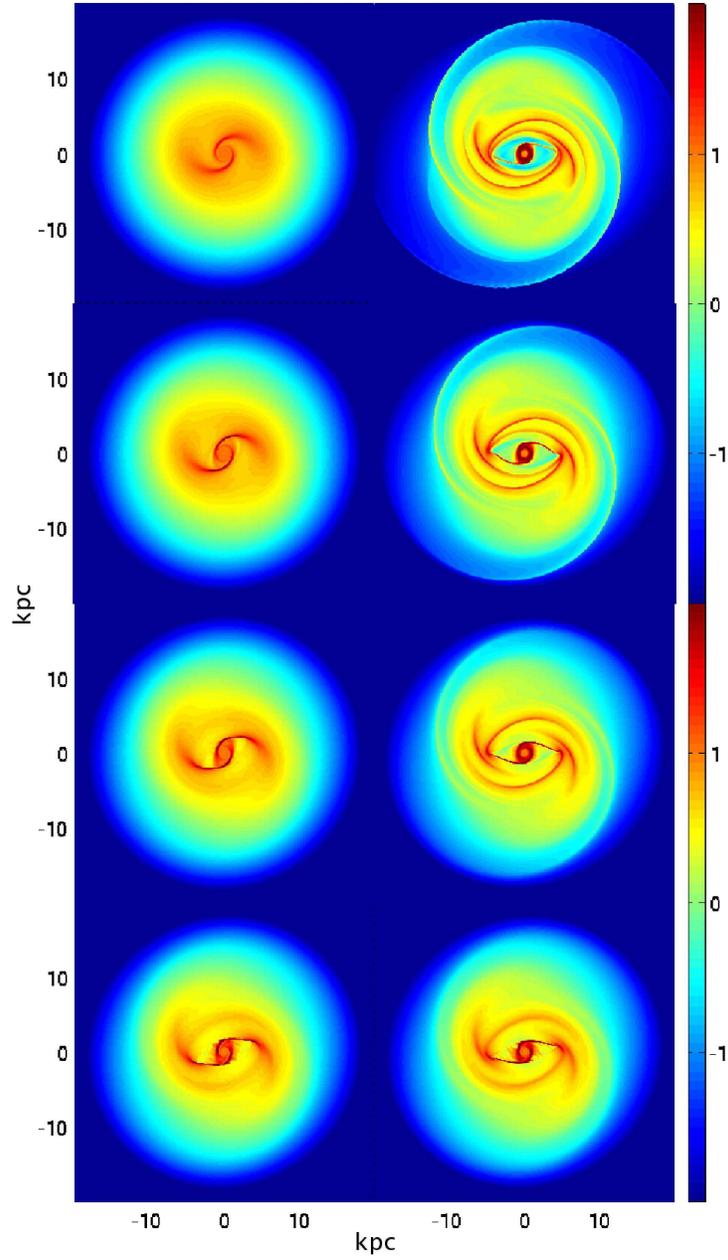}
\caption{Evolution of the bar-driven gaseous disk. The eight panels are arranged in counterclockwise 
order with increasing time. The individual times are 123, 154, 194, 237, 289, 341, 400, 532 Myr. The bar is in horizontal orientation in each frame. The color map denotes the surface density distribution in logarithmic scale.}.
\label{gasdisk}
\end{figure}

The evolution of the density features has been followed in the simulation to gain insight into their 
formation.  Initially, spiral waves excited at the ILR appear first.  As the strength of the bar 
potential is increased, the spiral waves intensify eventually leading to the formation of a shock. 
Subsequently, the inner parts of the spirals become straighter along the major axis of the bar and 
waves propagate inward.  Simultaneously, the waves excited at the inner 4:1 resonance appear, growing
stronger and propagating inward.  Their propagation leads to their interaction with the waves excited 
at the ILR to connect with the straight gas lanes  
to form the pointy oval inner ring.  This evolution is consistent with Schwarz's (1984) suggestion that 
inner rings are connected to the inner 4:1 resonance located just inside the CR.  The straight gas 
lanes propagating inward become more tightly wound and gradually form a nuclear ring.  A quasi-steady 
state is reached after one and three fourths revolutions of the bar, or 430 Myr.  The sequential 
images are shown in Figure~\ref{gasdisk}.

The general result concerning the formation of a nuclear ring is similar to that found by Piner, 
Stone \& Teuben (1995) in their numerical simulations of barred galaxies. The nuclear ring has a rich 
structure and resembles a tightly wrapped spiral.  This is consistent with the result presented 
here, but the requirements necessary for its formation differ in detail. Specifically, Piner et al. 
(1995) require that their model galaxy must have two inner Lindblad resonances and the extrema of the $\Omega - \kappa/2$ curves along the major and minor axes between 
the resonances must be at the same radial position.  In contrast, only one inner Lindblad resonance 
exists in our model galaxy.  As the density waves excited at the ILR propagate inward, their wavelength 
decreases, evolving to a more tightly wound structure about the center (Yuan \& Yang 2006).  As the 
galaxy evolves, mass is continuously accreted into the central region.  The tightly wound spirals 
become more dense and a ring is formed.

The existence of strong inner rings, offset dust lanes, and circumnuclear rings has also been 
found in the N-body simulation study by Byrd et al. (1994).  Some of their numerical results 
resemble NGC 6782 provided that the pattern speed of NGC 6782 is slow such that both an inner 
second harmonic resonance and an inner Lindblad resonance are allowed.

\begin{figure}
\figurenum{13}
\epsscale{1.0}
\plotone{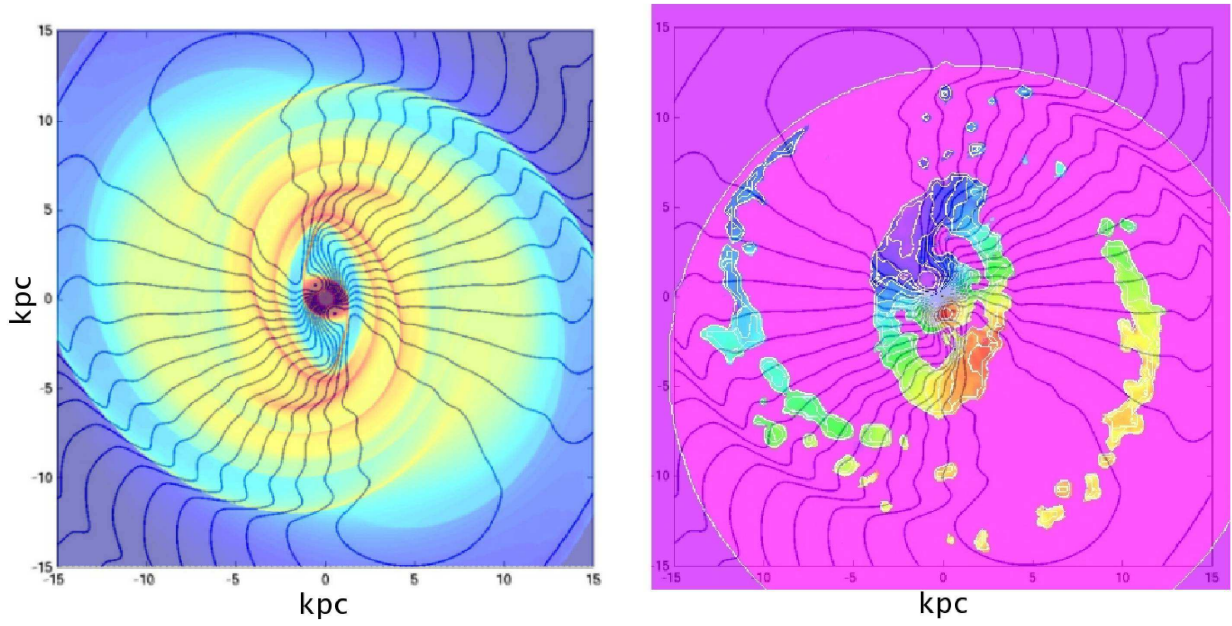}
\caption{Isovelocity curves from the simulation velocity field are plotted on top of the simulation density map (a)(left) and the observed Fabry-Perot velocity field (b)(right).}.\label{isovels}
\end{figure}

The isovelocity contours from the simulation velocity field are plotted on top of both the simulation 
density map and the observed Fabry-Perot velocity field in Figures~\ref{isovels}a and b respectively.  For these 
comparisons, we have smoothed the simulation velocity field to the same spatial resolution as that 
of the observation. Strong velocity gradients across the 
dust lanes are seen in Figure~\ref{isovels}a. 
Across the dust lanes, matter collides with the arm from the concave (or inner)
side and is deflected by the shock, resulting in mass inflow to fuel the nuclear star formation ring. 
It is noteworthy in Figure~\ref{isovels}b that at the southeastern part of the pointy inner ring, the velocity 
contours bend outwards along the spiral arms in both the simulation and observation.  This is 
one of the distinct features for the density waves excited at the ILR and is an important diagnostic.
Based upon this comparison, the inner ring is directly related to the ILR.  

The outward bending of the curves is caused by the same mechanism that produces the sharp velocity gradients across the dust lanes.
From the amount of bending of the isovelocity curves at the respective positions, 
it is inferred that the shock in the dust lanes is stronger than that in the pointy inner ring. 
Since the deflection of the isovelocity curves across the dust lanes is very large, the amount of 
shear is so high that it is likely that a cloud at the shock loci will be sheared out before collapsing. 
Therefore, the absence of HII regions or signs of recent star formation in these dust lanes can be 
naturally understood.  On the other hand, clouds located at the position of the density enhancements
in the inner ring, which undergo compression upon passage through the shock, will have sufficient 
time to collapse before being sheared out.  As a result, the inner ring shows clear signs of 
recent star formation.  This is consistent with Athanassoula's (1992) result, according to which 
these features should exist in strongly barred galaxies.  Although the ratio of the effective radial
force to the force for the circular motion is 5\% at OLR, it is 21\% at ILR.  Since almost all 
the important features of our simulation lie inside the CR, we conclude that the force associated 
with the rotating bar in our simulation for NGC 6782 is strong. For comparison, Laurikainen
et al. (2004) used a near-infrared image from the Ohio State University Bright Galaxy Survey
(Eskridge et al. 2002) to estimate the maximum relative torque in NGC 6782. Their value,
$Q_g$ = 0.165$\pm$0.008, located at r = 24.4 arcsecs (equal to 4.61 kpc), indicates only a moderate bar strength characteristic of
SAB galaxies (see also Buta \& Block 2001).
The value of $Q_g$ in our model for the best fit case, which is located at a different distance, r = 3.2 kpc, is found to be 0.135.

\subsection{HII Regions in the Pointy Inner Ring}

As noted in the paper by Crocker et al. (1996) and shown again in Figure~\ref{images}, the HII regions in the pointy inner ring 
of NGC 6782 are bright around the cusps and mostly concentrated in the vicinity of the ring major axis.
In their sample of 32 ringed galaxies, not only NGC 6782 but also IC 1438 and UGC 12646 exhibit most of 
their emission near the ring major axis with very little emission near the ring minor axis.  This is 
similar to NGC 1433 (Buta et al. 2001) and NGC 3081 (Buta et al. 2004), but there is more emission near the minor axes in these two cases.
An explanation for this phenomenon has been suggested by Contopoulos (1979; see also Byrd et al. 2006) involving the relative particle 
motions within the CR region.  In particular, particles would remain longer at the apocentra of their orbits 
than at the pericentra, which would enhance the density along the major axis of all orbits within the CR.
Thus, near the true major axis of an elongated inner ring which lies inside CR, HII regions are expected 
to be concentrated.

\begin{figure}
\figurenum{14}
\epsscale{1.0}
\plotone{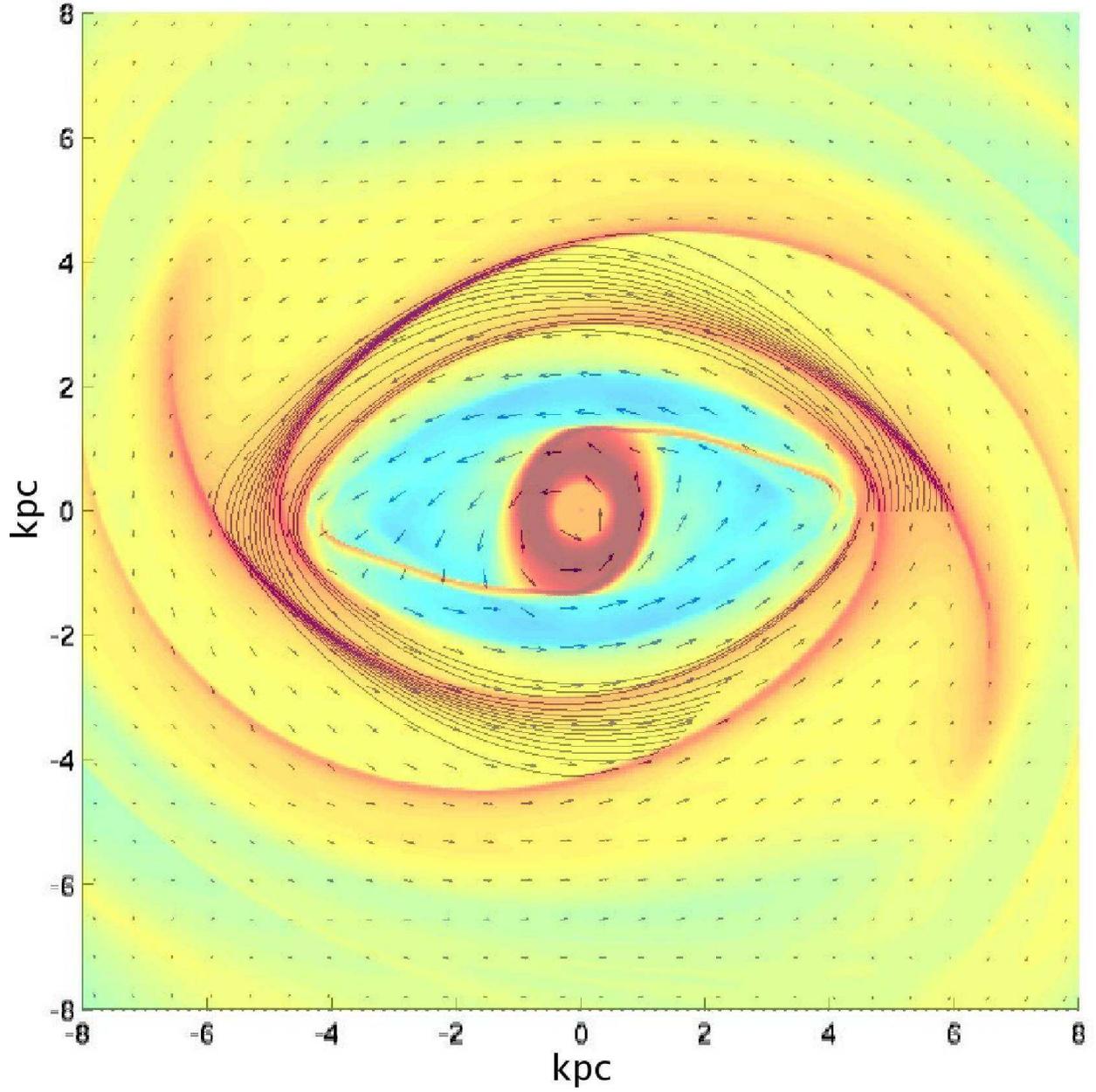}
\caption{Gas flow lines and velocity vectors in a reference frame in which the bar is 
at rest superposed on the surface density map.}.\label{gasflows}
\end{figure}

In our fully hydrodynamical simulation, the high density region around the two tips of the pointy inner ring is 
explained in terms of the interaction of the gas with the curved shocks.  To see this, we present the flow 
pattern in a reference frame rotating with the bar in Figure~\ref{gasflows}. Here, the arrows represent the velocity 
vectors at a given point overlaid on the face-on surface density map of the gas.  Since we seek to 
understand the existence and spatial distribution of the HII regions in the inner ring, the area in  
Figure~\ref{gasflows}
lies inside the CR, and we only plot the flow lines related to the inner ring. Note that since the gas 
flow field in our simulation is not steady, but only quasi-steady we use flow lines rather than 
streamlines to demonstrate their instantaneous directions.
In order to see clearly how flow lines bunch up, we only plot them for 
less than one revolution.  Flow lines start from a number of equidistant 
points along the direction of the bar semimajor axis along the x-axis.  Focusing on the left side of 
Figure~\ref{gasflows}, 
it can be seen that they hit the shock on the upper left and are deflected and bunched up to compress the 
gas there.  The compressed gas may trigger star formation and then move along the flow lines.  A moment later, 
the gas hits the shock on the lower left and the same process repeats itself.  From the velocity vectors, 
it can be seen that the gas around the tips of the pointy inner ring is moving more slowly than that near 
the ring minor axis, implying that the gas remains for a longer time near the ring major axis, consistent 
with Contopoulos' result (1979).  As a result, the HII regions in the pointy inner ring would be expected to be 
present around the cusps.

\subsection{Self-gravitation of the Gas Disk}

Simulations carried out with and without self-gravitation produce similar morphologies with the same 
overall shapes.  The primary quantitative differences are the radii of the nuclear rings. In particular, 
the nuclear ring in the simulation with self-gravitation is larger in extent than in the case without 
the inclusion of self-gravity effects.  This result is consistent with the asymptotic solutions in Yuan 
\& Cheng (1989) and Yuan \& Kuo (1997) who showed that the self-gravity will affect the radial location 
and orientation of the spiral pattern, but not its overall shape.  This result follows from the fact that 
self-gravitational effects will shift the Q-barrier to move toward the corotation region. As a result, the entire 
spiral pattern is shifted toward corotation, inward for the OLR and outward for the ILR. Since the 
spiral pattern in NGC 6782 lies inside the corotation point, the effect of self-gravity should move the spiral 
pattern outward and increase its extent.  For stronger self-gravity, the effect is greater and it would
be manifested most clearly on the dense nuclear ring in the simulation result. 

\section{Discussion}

Our findings on NGC 6782 may be compared with other recent studies. Patsis et al. (2003)
analyzed the shapes of inner SB rings from the point of view of two- and three-dimensional
periodic orbits in a potential consisting of a Miyamoto disk, a Plummer bulge, and a 3D Ferrers
bar, and considered the case of NGC 6782 as an example. They concluded that the pointed-oval
inner ring of NGC 6782 can be interpreted in terms of 2D 6:1 orbits of the ``$s$ family"
and two associated groups of stable 3D families. No orbits of the bar-supporting $x_1$
family were needed to explain the shape and location of the ring.

Rautiainen et al. (2005) used near-infrared images to infer gravitational potentials
in 38 galaxies, including NGC 6782,
and then to evolve a cloud-particle disk in these potentials until the
simulated morphology matches the $B$-band morphology. The goal was to measure the ratio
of the corotation radius ($R_{CR}$) to the bar radius ($R_{bar}$) 
and compare it with model expectations for
fast and slow bars. For NGC 6782, the best-fitting simulation gave a corotation radius
of 37\rlap{.}$^{\prime\prime}$1, compared to 47\rlap{.}$^{\prime\prime}$2 in our model.
With a measured bar radius of 31\rlap{.}$^{\prime\prime}$3, Rautiainen et al. derived
$R_{CR}/R_{bar}$=1.2, in the ``fast bar" domain of Debattista \& Sellwood (2000). For our
value of $R_{CR}$, the ratio is 1.5, which would be in the ``slow bar" domain. As we
have shown, NGC 6782 has an oval zone that almost fills the minor axis dimension of the
outer R$_1$ ring. Our CR radius fully encompasses this feature which extends well
beyond the ends of the main bar.

Quillen \& Frogel (1997) focussed on the shape of the outer R$_1$ ring in NGC 6782 as a means
of deducing the mass-to-light ratio of the bar. For our adopted orientation parameters,
we have shown that the deprojected position angle of the R$_1$ ring relative to the bar
is 74$^{\circ}$$\pm$3$^{\circ}$, while Buta (1995) derived a mean orientation of 
90$^{\circ}$$\pm$9$^{\circ}$ for R$_1$ rings. On the assumption that the ring is actually aligned
perpendicular to the bar, Quillen \& Frogel (1997) concluded that the inclination of
the galaxy must lie in the range 35$^{\circ}$ $<$ $i$ $<$ 45$^{\circ}$. To reproduce the
morphology of the ring for an inclination of 41$^{\circ}$, Quillen \& Frogel deduced that
the bar mass-to-light ratio must be greater than 70\% of that required for a maximal disk.
This study lacked the rotation information that we have available in our paper.

Romero-Gomez et al. (2006) have considered an alternative interpretation
of the rings in a galaxy like NGC 6782. They focussed on what they referred
to as "rR$_1$" rings, a combination of an inner ring and an R$_1$ outer ring
like that seen in NGC 6782. Instead of specific resonances, this pattern
is attributed to what the authors call ``invariant manifolds" of periodic
orbits associated with the L$_1$ and L$_2$ Lagrangian points in the potential
of a rotating bar. The manifolds can account for a highly elongated,
pointed-oval inner ring aligned with the bar and an R$_1$ outer ring which
dips in at its minor axis. Romero-Gomez et al. (2007) further show that
the variety of ring types, such as R$_1$R$_2^{\prime}$ and R$_2^{\prime}$,
as well as open spirals such as is seen in NGC 1365,
can also be explained by these manifolds.

\section{Conclusions}

We have performed two dimensional modeling of the barred spiral galaxy NGC 6782 using the Antares 
hydrodynamic code in the isothermal approximation within the framework 
of a simple model based on a nearly flat rotation curve (Eq. (9)), which is fit from the observations.
For a perturbing bar potential (Eqs. (6) and (7)) described by the angular speed, the location of the 
potential minimum, and the strength of the bar, we show that the primary density and velocity features 
of the galaxy can be reproduced provided that the parameters are in the range, 
$\Omega_p$ = 25 km s$^{-1}$ kpc$^{-1}$,
a$_1$ = 3.2 kpc, and B = 5\%.
Specifically, the observed shapes and positions of the bright nuclear ring, offset gas lanes, and the 
pointy oval inner ring are well reproduced in our model.  In addition, our simulation results 
also reproduce the two faint branches near the tip of the inner ring at the northwestern outer arm in 
the $B$-band image.

Based on the time-dependent evolution of the gas disk, the formation of the primary features has been 
examined.  In particular, it has been found that the pointy inner ring is formed via the interaction 
between the density waves excited at the ILR and the 4:1 higher harmonic density waves under the 
gravitational influence of a slowly rotating strong bar.  The nuclear ring, on the other hand, is 
initially a pair of tightly wound spirals which gradually evolve into a ring after matter has continuously 
accreted into the central region.

Upon comparison with the observed velocity field, the numerical simulations provide an understanding for 
the origin of the outward bending of the isovelocity contours along the spiral arms at the southeastern part of the pointy inner ring.
This feature is caused by gas colliding with the spiral arm from the interior (concave side). Finally, 
the existence of cuspy HII regions in the pointy oval inner ring can also be naturally explained by the 
concentration of gas flow lines around the tips of the inner ring.
  
In the future, high resolution observations are desirable in order to discern the strong velocity gradients 
across the dust lanes seen in our simulation to provide further confirmation of the model presented in 
this paper.

\acknowledgments
We thank Ted Williams for assistance with the Fabry-Perot observations of NGC 6782,
and we also thank Guy B. Purcell and Ted Williams for the programs used for reducing
the data. 
We are grateful to Ronald E. Taam and Paul T.P. Ho for many valuable comments. 
R.B. acknowledges the support of NSF Grant AST-0507140 to the University of Alabama.
L.-H.L. and C.Y. acknowledge the support of National Science Council, Taiwan, NSC96-2752-M-001-002-PAE.

\clearpage

\begin{deluxetable}{ll}
\tabletypesize{\scriptsize}
\tablewidth{0pc}
\tablecaption{Global Optical Characteristics of NGC 6782}
\tablehead{
\colhead{Parameter} &
\colhead{Value} 
}
\startdata
$\alpha(1950)$\tablenotemark{a} & 19$^h$ 19$^m$ 35\rlap{.}$^s$1\\
$\delta(1950)$\tablenotemark{a} & $-$60$^{\circ}$ 01$^{\prime}$ 12$^{\prime\prime}$\\
SGC type\tablenotemark{b} & (R)SAB(r)a \\
CSRG type\tablenotemark{c} & (R$_1^{\prime}$)SB(r)0/a \\
de Vaucouleurs Atlas type\tablenotemark{d} & (R$_1$R$_2^{\prime}$)SB(r)a \\
logD$_{25}$ (0\rlap{.}$^{\prime}$1)        &  1.355 \\
logR$_{25}$        &   0.050 \\
$\phi_{25}$        &   52\rlap{.}$^{\circ}$5 (1950) \\
$B_T$              &   12.85  \\
$(B-V)_T$          &    0.91  \\
$(V-I)_T$          &    1.22  \\
Photometric axis ratio $<q_p>$ (70$^{\prime\prime}$ $\leq$ $r$ $\leq$ 89$^{\prime\prime}$) & 0.894 $\pm$ 0.002 \\
Photometric major axis $<\phi_p>$ (70$^{\prime\prime}$ $\leq$ $r$ $\leq$ 89$^{\prime\prime}$) & 34\rlap{.}$^{\circ}$9 $\pm$ 0\rlap{.}$^{\circ}$5 (1950) \\
Photometric inclination $i_p$\tablenotemark{e}                &    27\rlap{.}$^{\circ}$2$\pm$0\rlap{.}$^{\circ}$2  \\
Kinematic inclination $i_k$                                        & 27$^{\circ}$$\pm$5$^{\circ}$ \\
Kinematic line of nodes position angle $\phi_k$\tablenotemark{f}                  & 35$^{\circ}$$\pm$2$^{\circ}$\\
Heliocentric systemic radial velocity $V_s$       & 3921$\pm$3 km s$^{-1}$ \\
Deprojected nuclear ring major axis diameter   & 9\rlap{.}$^{\prime\prime}$9$\pm$0\rlap{.}$^{\prime\prime}$2 \\
Deprojected nuclear ring axis ratio            & 0.94$\pm$0.02  \\
Deprojected inner ring major axis diameter   & 53\rlap{.}$^{\prime\prime}$2$\pm$0\rlap{.}$^{\prime\prime}$2 \\
Deprojected inner ring axis ratio            & 0.687$\pm$0.004  \\
Deprojected bar-inner ring position angle    & 1\rlap{.}$^{\circ}$8$\pm$1\rlap{.}$^{\circ}$1  \\
Deprojected outer R$_1$ ring major axis diameter   & 130$^{\prime\prime}$$\pm$1$^{\prime\prime}$ \\
Deprojected outer R$_1$ ring axis ratio            & 0.79$\pm$0.01  \\
Deprojected bar-outer R$_1$ ring position angle    & 74$^{\circ}$$\pm$3$^{\circ}$  \\
\enddata
\tablenotetext{a}{de Vaucouleurs et al. (1991)}
\tablenotetext{b}{Corwin et al. (1985)}
\tablenotetext{c}{Buta (1995)}
\tablenotetext{d}{Buta et al. (2007)}
\tablenotetext{e}{Based on $<q_p>$ and assuming the outer isophotes represent an oblate spheroid having an intrinsic flattening of 0.2}
\tablenotetext{f}{Assuming pure circular motion}
\end{deluxetable}

\clearpage

\end{document}